\documentclass[twocolumn,groupedaddress,nofootinbib, superscriptaddress]{revtex4}
\usepackage{amsmath,amssymb,amsfonts}
\usepackage{graphicx}
\usepackage{graphicx}
\usepackage{setspace}

\begin{document}

\title{Controlled formation and reflection of a bright solitary matter-wave }

\author{A. L. Marchant}
\email{a.l.marchant@durham.ac.uk}
\affiliation{Joint Quantum Centre (JQC), Durham - Newcastle, Department of Physics, Durham University, Durham DH1 3LE, United Kingdom}
\author{T. P. Billam}
\affiliation{Jack Dodd Centre for Quantum Technology, Department of Physics, University of Otago, Dunedin 9016, New Zealand}
\author{T. P. Wiles}
\affiliation{Joint Quantum Centre (JQC), Durham - Newcastle, Department of Physics, Durham University, Durham DH1 3LE, United Kingdom}
\author{M. M. H. Yu}
\affiliation{Joint Quantum Centre (JQC), Durham - Newcastle, Department of Physics, Durham University, Durham DH1 3LE, United Kingdom}
\author{S. A. Gardiner}
\affiliation{Joint Quantum Centre (JQC), Durham - Newcastle, Department of Physics, Durham University, Durham DH1 3LE, United Kingdom}
\author{S. L. Cornish}
\affiliation{Joint Quantum Centre (JQC), Durham - Newcastle, Department of Physics, Durham University, Durham DH1 3LE, United Kingdom}

\date{\today}

%\pacs{
%03.75.Lm,     	% Solitons
%03.75.-b,			% Matter waves
%05.45.Yv,			% Solitons
%03.75.Kk			% Dynamical properties of BEC
%}

\maketitle

%----------------------------------------------------------------------------------------------------------------------------
%----------------------------------------------------- INTRODUCTION ---------------------------------------------------------

\textbf{Solitons are non-dispersive wave solutions that arise in a diverse range of nonlinear systems, stablised by a focussing or defocussing nonlinearity. First observed in shallow water \cite{JSR.solitons}, solitons have subsequently been studied in many other fields including nonlinear optics, biophysics, astrophysics, plasma and particle physics \cite{Phys.Solitons}. They are characterised by well localised wavepackets that maintain their initial shape and amplitude for all time, even following collisions with other solitons. Here we report the controlled formation of bright solitary matter-waves, the 3D analog to solitons, from Bose-Einstein condensates of $^{85}$Rb and observe their propagation in an optical waveguide. These results pave the way for new experimental studies of bright solitary matter-wave dynamics to elucidate the wealth of existing theoretical work and to explore an array of potential applications including novel interferometric devices \cite{RevModPhys.81.1051}, the study of short-range atom-surface potentials \cite{CornishPhysicaD} and the realisation of Schr\"odinger-cat states \cite{PhysRevLett.102.010403, PhysRevA.80.043616}. } 

Bose-Einstein condensates formed from dilute atomic gases support bright soliton solutions for attractive interatomic interactions (focussing nonlinearity), manifesting themselves as localized humps in the field amplitude. In contrast, dark solitons appear as localized reductions in an otherwise uniform field amplitude, preserved by a defocussing nonlinearity (repulsive interactions). The control with which these systems can be manipulated, combined with the unique properties of matter-wave solitons, leads to a rich testing ground for theoretical descriptions of quantum many-body systems. Condensates are commonly described by a mean-field treatment \cite{PandS, PitaevskiiandS} leading to the well-known Gross-Pitaevskii equation (GPE) in which the atomic interactions are described by a nonlinear term proportional to the $s$-wave scattering length $a_s$ and the condensate density. In the one-dimensional (1D), homogeneous limit the GPE takes the form of a nonlinear Schr\"odinger equation which supports a spectrum of mathematically ideal soliton solutions. Experiments approach this theoretically ideal scenario by confining the condensate in an elongated, prolate trap typically with tight radial confinement. However, this quasi-1D geometry is usually accompanied by the presence of weak axial harmonic trapping which removes the integrability of the system and prevents the appearance of true solitons. Nevertheless, solitary wave solutions remain which retain many similarities to the classical soliton solutions \cite{Bookchapter}, such as propagation without dispersion. 

Previously, bright solitary matter-waves have been realised in three separate experiments \cite{2002Natur.417..150S, Khaykovich17052002, PhysRevLett.96.170401}. In each case a Feshbach resonance was used to switch the interactions from repulsive ($a_s>0$) to attractive ($a_s<0$) in order to form solitary waves out of the collapse instability \cite{PhysRevA.51.4704}. In two of these experiments, multiple wavepackets were created, allowing the study of the dynamics during collisions in the trap. The observation of solitary waves raises many interesting questions regarding the relationship between the mathematical ideal and the experimental reality. It is unclear how soliton-like the solitary waves created in experiments with finite radial trapping and harmonic axial confinement are.  An answer to this question needs to be established before potential applications utilising solitary waves can be realised. At a more fundamental level it remains to be tested whether or not the GPE treatment fully describes the solitary waves created in experiments. Solitary waves realised experimentally typically contain $\lesssim$1000 atoms, placing them well outside of the thermodynamic limit and potentially outside the reach of the mean-field description. Several theoretical studies of bright solitary waves beyond the mean-field description have now been performed, either including effects of quantum noise using the truncated Wigner method \cite{Davis2009} or using approximate analytic and numerical methods to simulate the full many-body problem \cite{PhysRevLett.100.130401, PhysRevLett.102.010403}. These generate results potentially in conflict with the behaviour predicted by the GPE treatment. 

In this work we report the controlled formation of bright solitary matter-waves from a $^{85}$Rb Bose-Einstein condensate. The experimental geometry is such that the velocity of the wavepackets can be precisely controlled, a key factor in facilitating the future exploration of solitary wave interactions and collisions. In addition, we observe and model the controlled reflection of solitary waves from a broad Gaussian potential barrier, demonstrating their particle-like nature. 
\begin{figure*}
	\centering
			\includegraphics[width=0.8\textwidth]{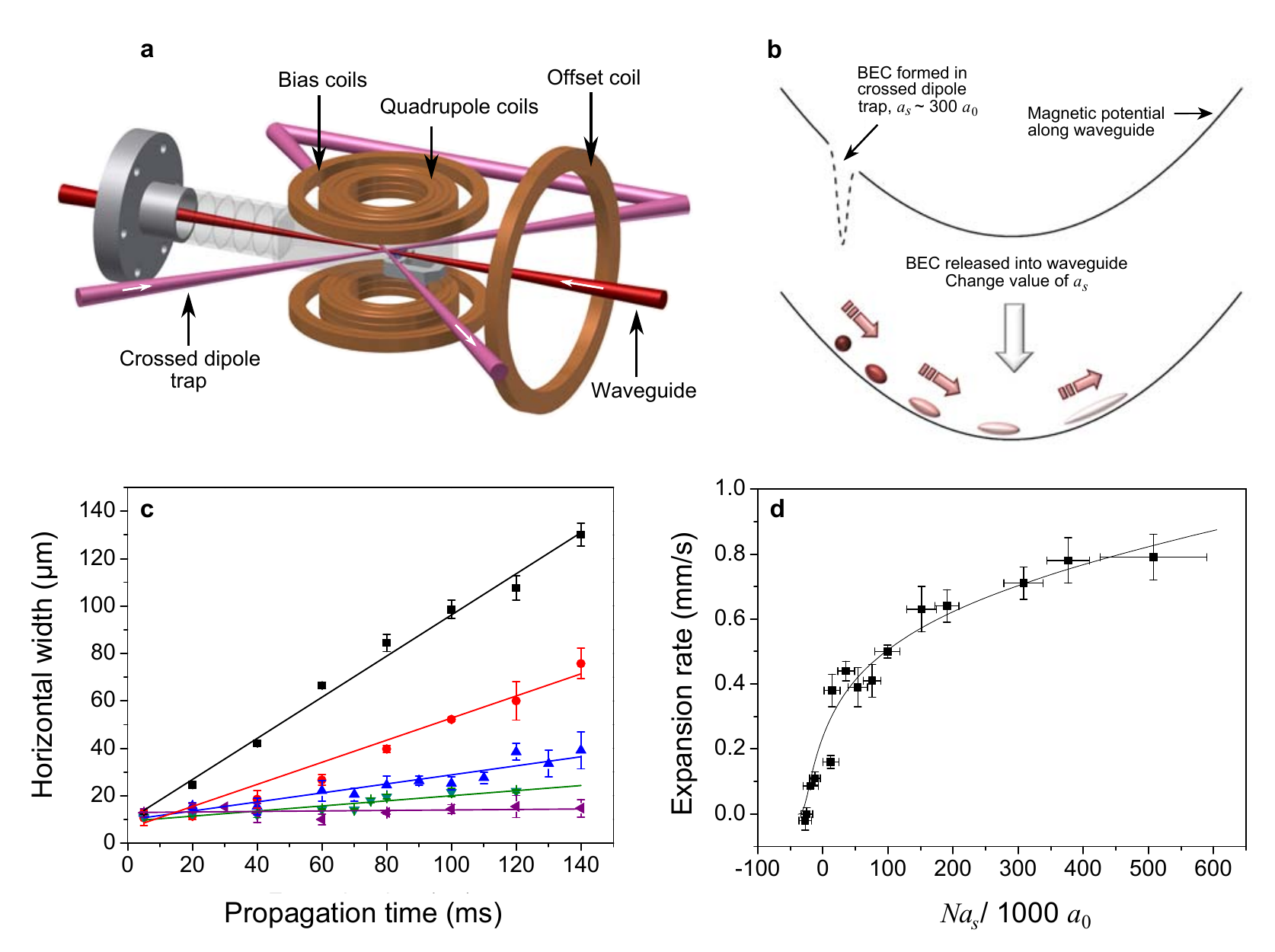}
	\caption{Expansion in the waveguide: (a) Experimental setup showing the glass science cell, the crossed dipole trap used to create the BEC, the optical waveguide and the quadrupole, bias and offset coils. Also shown in the cell is a super-polished Dove prism (blue), mounted on a macor support, to be used for future experiments. (b) Schematic of the release of the condensate from the crossed dipole trap into the waveguide. (c) Condensate expansion in the waveguide for $a_s=$ 165~$a_0$ (black), 23~$a_0$ (red), 4~$a_0$ (blue), -7~$a_0$ (green) and -11~$a_0$ (purple). Solid lines are linear fits to the experimental data where the widths are rms values. (d) Condensate expansion rate in the waveguide as a function of atom number and scattering length. The solid line is the theoretical expansion rate calculated from a zero-temperature simulation of the experimental expansion using a cylindrically-symmetric, 3D Gross-Pitaevskii equation. As in the data, the expansion rate is defined using the change in the width of the BEC between 10~ms and 100~ms after release into the waveguide potential, which is approximately linear over this time interval in all cases}
	\label{fig:Expansion}
	\end{figure*}

%----------------------------------------------------------------------------------------------------------------------------
%---------------------------------------------- Propagation in Waveguide -------------------------------------------------------

$^{85}$Rb is a prime candidate for solitary wave experiments owing to the existence of a broad Feshbach resonance at $\sim$155~G in collisions between  atoms in the $F=2, m_F=-2$ state. We use this resonance to form a stable, repulsively interacting condensate in a crossed optical dipole trap, shown in Fig.~\ref{fig:Expansion}(a). The condensate is then loaded into a quasi-1D waveguide, better suited geometrically to the observation of solitary waves. At the point of release into the waveguide, the magnetic bias field controlling the atomic scattering length is jumped to a new value (see Fig.~\ref{fig:Expansion}(b)). As the BEC propagates along the waveguide the value of $a_s$ determines the rate of expansion of the condensate in the axial direction. We probe this expansion by measuring the condensate size as a function of time for different values of $a_s$ as shown in Fig.~\ref{fig:Expansion}(c). Fitting the experimental data we can extract an expansion rate for the BEC, dependent on $a_s$ and $N$. This is shown in Fig.~\ref{fig:Expansion}(d), along with a 3D GPE simulation of the expansion (the solid line). At $a_s=-11~a_0$ and $N=2,000$ we see the expansion rate of the BEC becomes consistent with zero. This lack of dispersion with time indicates the formation of a bright solitary matter-wave. 
\begin{figure*}
	\centering
		\includegraphics[width=0.9\textwidth]{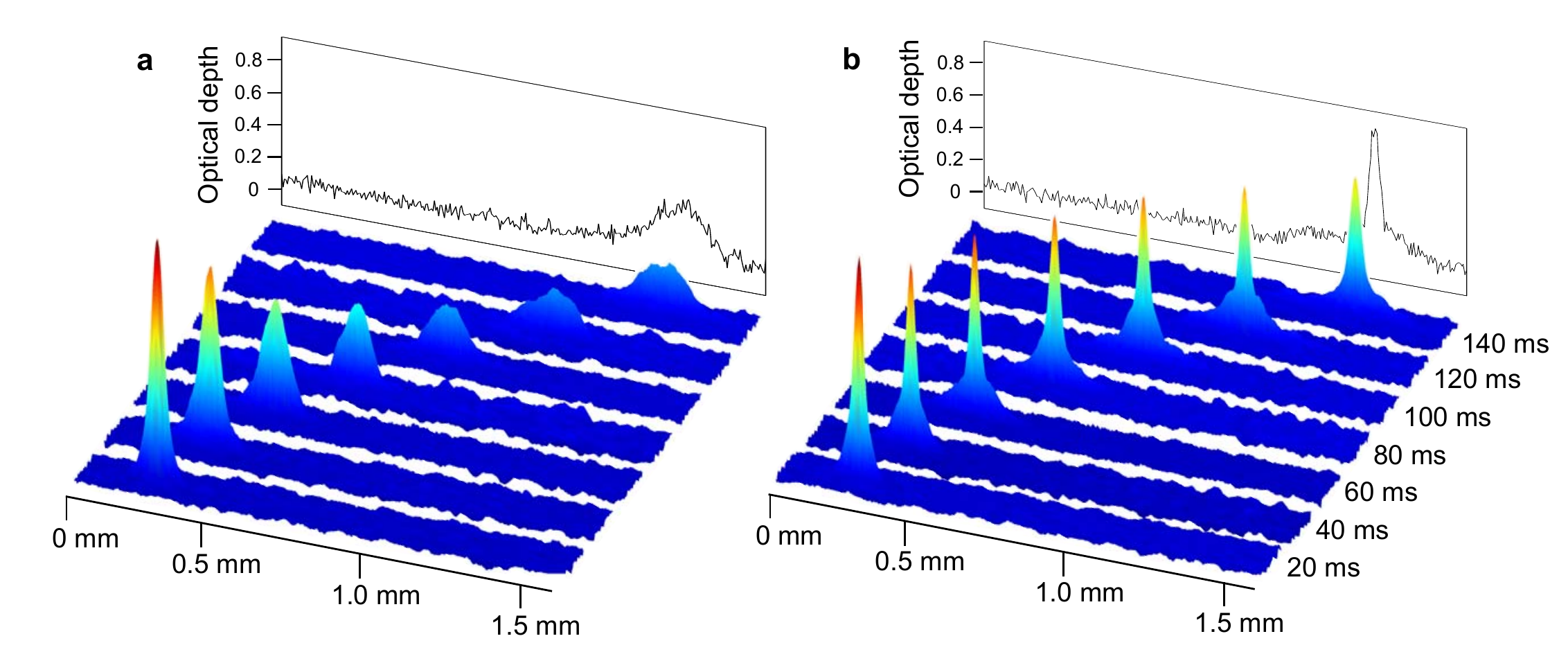}
	\caption{Propagation in the waveguide: (a) As a repulsive BEC propagates along the waveguide the atomic interactions cause the condensate to spread, leading to a drop in optical depth. (b) In contrast, the attractive interactions present in a bright solitary matter-wave hold the atomic wavepacket together as it propagates, maintaining its shape with time. Crosscuts shown are the horizontal optical depth profiles of the condensates after 140~ms propagation time along the waveguide. }
	\label{fig:Propagation3D}
\end{figure*}

Figure~\ref{fig:Propagation3D} shows the propagation of this solitary wave, contrasted to that of a repulsively interacting BEC. As the repulsive wavepacket propagates the axial expansion causes a significant drop in optical depth not seen for the solitary wave. We observe a solitary wave propagating over a distance of 1.1~mm in a time of $\sim$150~ms with very little distortion. 

%----------------------------------------------------------------------------------------------------------------------------
%---------------------------------------------- Reflection from 532nm barrier -----------------------------------------------

\begin{figure*}
	\centering
		\includegraphics[width=0.9\textwidth]{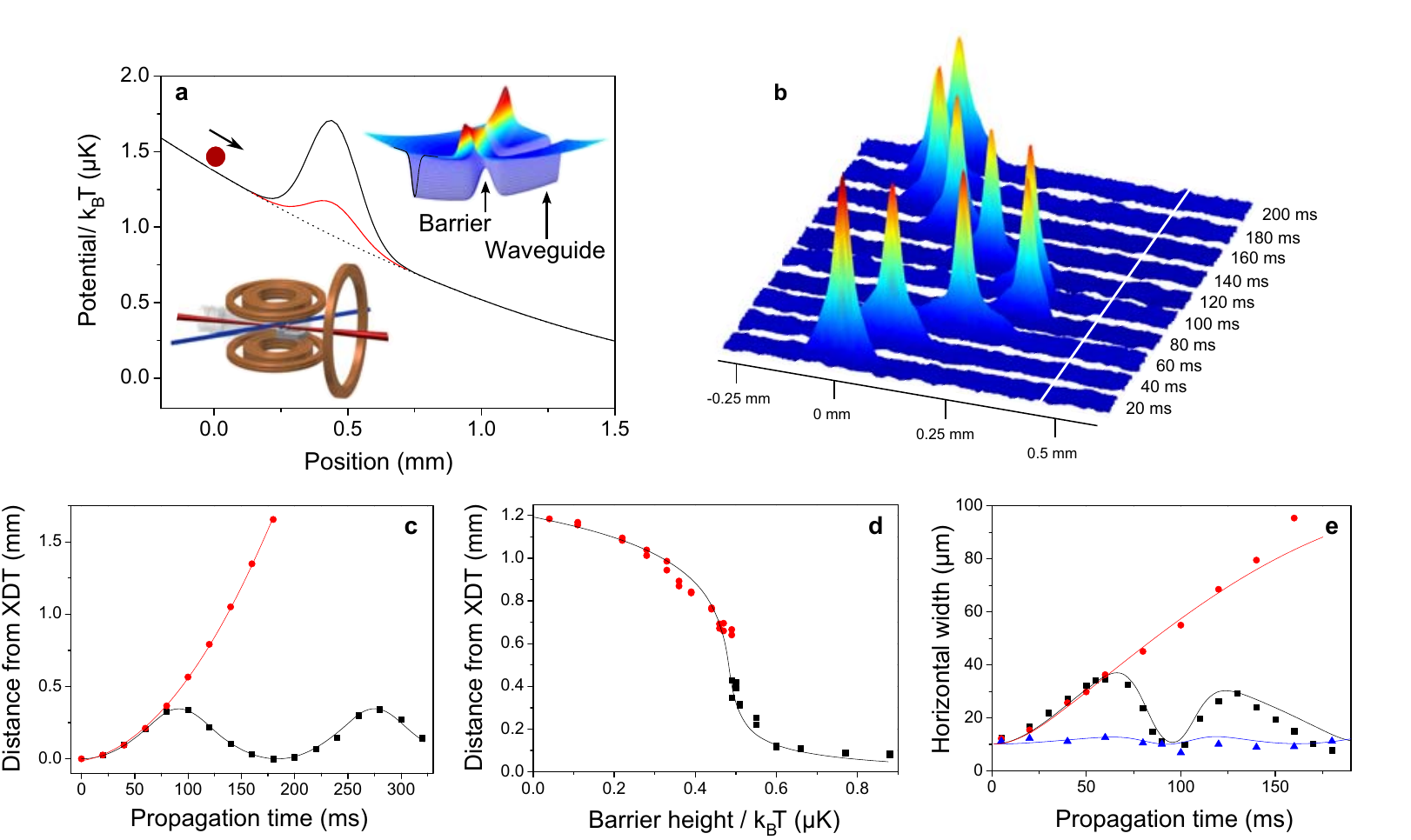}
	\caption{Reflection from a repulsive Gaussian barrier: (a) Potential in the axial direction along the waveguide in the presence of the repulsive barrier. (Inset, upper: Combined waveguide and Gaussian barrier potential. Lower: Experimental setup.) (b) False colour images of a solitary wave reflecting from the barrier. The white line shows the location of the barrier centre. (c) Horizontal position of a solitary wave propagating in the waveguide in the absence (red) and presence (black) of the repulsive barrier. (d) The position of a solitary wave after 150~ms propagation time as a function of the barrier height. Red (black) points correspond to the solitary wave travelling over (being reflected from) the barrier. Solid lines in (c) and (d): Theoretical trajectory calculated using a classical particle model with no free parameters. (e) Condensate width following reflection from the barrier. In the absence of a barrier, a repulsive BEC will expand as it propagates (red). With the barrier in place, an oscillation in the condensate width is set up following the strong compression of the condensate at the barrier due to the shape of the potential (black). A solitary wave undergoing the same collision emerges unaltered (blue). Solid lines are the theoretical condensate widths calculated by solving the 3D (cylindrically symmetric) GPE.}
	\label{fig:Barrier_combined}
\end{figure*}

To probe the stability of the solitary wave we investigate reflection of the wavepacket from a repulsive Gaussian barrier with a $1/e^2$ radius of 130~$\mu$m, shown in Fig.~\ref{fig:Barrier_combined}(a). Figures \ref{fig:Barrier_combined}(b) and (c) show the position of the solitary wave as a function of time in the presence of a 760~nK barrier potential. In this case the barrier height is greater than the kinetic energy of the solitary wave and the wavepacket is cleanly reflected.    

Using a barrier much wider than the solitary wave size the atomic center-of-mass coordinate behaves classically, with the solitary wave acting as a single particle 'rolling up a potential hill'. By varying the height of the potential barrier it is possible to select whether the solitary wave is reflected or allowed to travel over the barrier. The position of the solitary wave after 150~ms is shown in Fig. \ref{fig:Barrier_combined}(d) as a function of barrier height. The solid line is a theoretical trajectory, calculated using classical mechanics with no free parameters, showing excellent agreement with the data.  

In Fig.~\ref{fig:Barrier_combined}(e) we compare the effect of reflection from the barrier for a solitary wave and a repulsive BEC and contrast the change in width to the case of a repulsive BEC propagating along the waveguide in the absence of the barrier. The solid lines are the theoretical condensate widths calculated by solving the 3D (cylindrically symmetric) GPE. As expected, the solitary wave is robust against collisions with a repulsive Gaussian barrier and following the reflection maintains its shape, continuing to propagate without dispersion. In the absence of the barrier the repulsive BEC expands steadily in time. (We attribute the disagreement between experiment and theory at longer times to a small thermal component making the measurement of the condensate width less accurate.) In the barrier reflection case, an oscillation in the condensate width is induced as a result of the larger spatial extent of the repulsive BEC causing it to be strongly compressed as it is reflected from the barrier. Such contrast in the behaviour of the repulsive BEC and the solitary wave reflection lends weight to previous theoretical prediction regarding the superior characteristics of solitary waves for observing quantum reflection from surfaces \cite{CornishPhysicaD}. 

%----------------------------------------------------------------------------------------------------------------------------
%------------------------------------------------------ Future work ---------------------------------------------------------

There is currently much theoretical interest \cite{PhysRevA.85.053621, PhysRevA.83.041602, 1367-2630-14-4-043040, PhysRevA.81.033614, Hulet_theory} in the scattering of solitary waves from narrow potential barriers where, if the barrier width is on the order of the solitary wave width, quantum effects are observable. At high kinetic energy soliton splitting is energetically allowed at narrow repulsive barriers. The effect of quantum tunnelling means the barrier can act as a beam splitter \cite{PhysRevA.85.053621}, dividing the soliton into two parts \cite{Huletexpt}. These multiple wavepackets can then be used to investigate the phase dependence of binary collisions \cite{PhysRevA.83.041602}, the behaviour of collisions of two solitary waves on a barrier \cite{1367-2630-14-4-043040, PhysRevA.85.053621} and would provide a solid first step towards the realisation of a bright solitary wave interferometer. In the limit of low kinetic energy a mean-field GPE treatment of the problem begins to break down \cite{PhysRevA.86.033608} and quantum behaviour, described by the Lieb-Liniger Hamiltonian \cite{PhysRev.130.1605}, becomes more significant. Here, splitting of the soliton is energetically forbidden and it becomes possible to create Schr\"odinger-cat states \cite{PhysRevLett.102.010403, PhysRevA.80.043616}.

The use of a narrow potential to controllably split a solitary wave presents an opportunity to investigate one of the key open questions arising from previous work; what governs the dynamics and stability of multiple solitary waves existing in the same trap? The long lived nature of the solitary waves and their apparent stability during binary collisions has been the subject of a wealth of theoretical work \cite{PhysRevLett.89.200404, Parker20091456, Parker.JPhysB, PhysRevLett.92.040401, Davis2009}. Within the framework of the GPE, the observed stability of soliton collisions can only be explained by imposing a relative phase $\phi=\pi$ between neighbouring solitary waves \cite{Parker.JPhysB} such that the collisions are effectively repulsive in character. Several other studies address the apparent stability of solitary waves in binary collisions, offering different interpretations which do not require the imposition of a relative phase $\phi=\pi$ between neighbouring solitary waves. The inclusion of quantum noise \cite{Davis2009} or accounting for many body effects \cite{PhysRevLett.100.130401} both result in effectively repulsive interactions between solitary waves, irrespective of initial phase. Interestingly, incoherent, fragmented objects are also predicted to form in the many body formalism. Further experimental studies are undoubtedly required to address the role of the relative phase in solitary wave collisions and to test the different theoretical descriptions of quantum many-body systems. 
   
Although reflection and splitting experiments show the potential to settle the theoretical debate over the solitary wave formation and dynamics, the ability to probe such narrow and hence rapidly varying potentials using these wavepackets also lends itself to an obvious application in precision measurement. Atoms close to a surface are subject to the short-range Casimir-Polder and van der Waals potentials which can be measured using the classical and quantum reflection of bright solitary matter-waves \cite{CornishPhysicaD}. Our apparatus includes a super-polished Dove prism for such studies, see Fig.~\ref{fig:Expansion}(a). Further in the future, the ability to deliver and manipulate ultracold atoms near to a solid surface may open up new routes to probe short range corrections to gravity \cite{PhysRevD.68.124021} due to exotic forces beyond the Standard model.

%------------------------------------------------- METHODS ---------------------------------
%------------------------------------------------- METHODS ---------------------------------

\section{Methods} 

\subsection{Production of a tunable Bose-Einstein condensate}

We create a Bose-Einstein condensate with tunable atomic interactions using the method described in \cite{PhysRevA.85.053647}. A magnetic Feshbach resonance is used to tune both the elastic and inelastic scattering properties of the atomic sample to achieve efficient evaporation. Importantly the resonance at 155~G in collisions between $^{85}$Rb atoms in the $F=2, m_F=-2$ state gives control over the $s$-wave scattering length close to the zero crossing of $\sim40~a_0/$G. 

The use of a magnetic Feshbach resonance means it is advantageous to work with a levitated crossed optical dipole trap. This is formed from a single 10.1~W, $\lambda=$1064~nm laser beam (IPG: YLR-15-1064-LP-SF) used in a bow-tie configuration as shown in Fig.~\ref{fig:Expansion}(a). The term `levitated' refers to the use of an additional magnetic quadrupole field whose vertical gradient is set to just less than that required to support atoms against gravity. This trap allows the magnetic field, and hence scattering length, to be changed independently of the trapping frequencies. 

\subsection{Loading the optical waveguide}

To investigate the creation of solitary waves we begin by forming a BEC containing up to 10,000 atoms at a scattering length of $a_{s}\approx300~a_0$. The crossed beam trap in which the BEC is created has a roughly spherically symmetric geometry at the point of condensation, with final trap frequencies of $\omega_{x,y,z}=2\pi\times(31,27,25)$~Hz. This trap is ill-suited to the observation of bright solitary matter-waves and thus we transfer the condensate into a more quasi-1D waveguide created by an additional 1064~nm laser beam, focused to a waist of 117~$\mu$m and intersecting the crossed trap at 45$^\circ$ to each beam. This enters the glass science cell through the back surface of an anti-reflection coated fused silica Dove prism (to be later used for the study of atom-surface interactions \cite{CornishPhysicaD}).

To load the condensate into the waveguide the scattering length is ramped close to $a_s=0$ in 50~ms thus reducing the condensate size and creating a BEC approximately in the harmonic oscillator ground state of the crossed trap. The BEC is then held for 10~ms to allow the magnetic field to stabilise before simultaneously switching the waveguide beam on, the crossed beams off and jumping the quadrupole gradient in the vertical direction from $B'=21.5$~G~cm$^{-1}$ to 26~G~cm$^{-1}$. Although it is advantageous in terms of the evaporation to be under levitated during the condensation phase, we must increase the gradient once we wish to transfer the atoms. This ensures a truer levitation of the atoms in the waveguide trap, thus maximising the trap depth of the beam. In addition, the presence of the quadrupole gradient provides much of the, albeit weak, axial trapping along the beam,
$\omega_{\rm{axial}}=1/2\sqrt{\mu B'^2/m B_0}\approx2\pi\times$1~Hz \cite{PhysRevA.79.063631}. Here, $\mu$ is the magnetic moment of atoms with mass $m$ and $B_0$ is the magnetic bias field. The waveguide beam itself contributes $<0.1$~Hz to the axial trapping, hence the magnetic confinement dominates in this direction. At a beam power of 0.17~W the waveguide and quadrupole potential produce a trap of $\omega_{x,y,z}=2\pi\times(1,27,27)$~Hz. Here the radial trap frequency ($\omega_{y,z}$) approximately matches that of the crossed beam trap at the point of condensation. 

\subsection{Propagation in the waveguide}

A small offset (2.6~mm) between the crossed dipole trap, i.e. the waveguide loading position, and the quadrupole centre means that once loaded into the waveguide, the BEC propagates freely towards the magnetic field minimum along the direction of the waveguide, undergoing harmonic motion. As the BEC propagates its rate of expansion in the axial direction is determined by the scattering length. Although strictly speaking the expansion is non-linear over the full range of times measured, a linear approximation is valid over the range $10$~ms$<t<$100~ms from which we can extract a `rate'.

%--------------------------------------------- Velocity control ---------------------------------------

\subsection{Control of the solitary wave velocity}

The position of the magnetic field zero in the axial direction of the waveguide can be displaced by an amount determined by the magnetic field gradient in this direction, $B'/2$, and a moderate offset field, $B_{\rm{offset}}$, according to $\Delta x=B_{\rm{offset}}/(B'/2)$ \cite{1367-2630-13-12-125003}. In this way the amplitude, and hence velocity, of the solitary wave motion can be precisely controlled due to the dominance of the magnetic potential over the optical confinement of the waveguide along the axial direction. The maximum velocity is given by $v=A\omega_{\rm{axial}}$ where $A$ is the amplitude of the motion, set by the separation between the minimum of the magnetic potential along the axis of the waveguide and the release point from the crossed dipole trap. Using this technique the solitary wave can reach velocities of tens of mm~s$^{-1}$ when travelling through the centre of the harmonic potential or, alternatively, be brought to a near standstill, achieving velocities $<0.5$~mm~s$^{-1}$.

%----------------------------------------- Classical reflection -----------------------------------

\subsection{Classical reflection from a Gaussian barrier}

To produce the repulsive potential barrier we use a 532~nm Gaussian laser beam (derived from a Laser Quantum Finesse laser), focussed to a waist of 131~$\mu$m horizontally and 495~$\mu$m vertically, with a power of up to 2~W. The barrier is aligned to cross the waveguide in the horizontal plane at an angle of $\sim$45$^\circ$ and is offset by 455~$\mu$m from where the BEC is released from the crossed dipole trap, see Fig.~\ref{fig:Barrier_combined}(a). This angle is restricted by the available optical access close to the trap centre.  

%----------------------------------------- Modelling -----------------------------------

\subsection{Theoretical modelling}

The release of the BEC into the waveguide potential, and its subsequent expansion, was modelled at zero-temperature by solving the Gross-Pitaevskii equation in 3D using a cylindrically-symmetric Fourier pseudospectral method. In all cases the initial non-interacting ground state of a harmonic trap with axial (radial) frequency 30~(27)~Hz (corresponding to the crossed dipole trap potential) was released instantaneously into another harmonic trap with axial (radial) frequency 1~(27)~Hz and offset by 2.6~mm along the axial direction (corresponding to the waveguide potential). The scattering length was instantaneously changed to the appropriate value of $a_s$ at the time of release. 

In cases where the barrier was present this was modelled as a Gaussian `light-sheet' potential centered on a plane perpendicular to the axial direction, offset from the initial harmonic trap by 2.145~mm, and with height 760~nK and width 131$\sqrt{2}~\mu$m. Compared to the experimental barrier beam this model neglects the vertical width of the beam which is large compared to the radial extent of the BEC in the waveguide, and includes the geometric factor $\sqrt{2}$ to account for the 45 degree angle of the beam.

Expansion rates were calculated from the full-width at half maximum of the BEC axial density profile predicted by the GPE (obtained by integrating over the radial coordinate) after 10~ms and 100~ms of expansion. In all cases, the change in radius over this time interval was approximately linear.  For the simulations in Fig.~\ref{fig:Barrier_combined}(e), the width was calculated by convolving the BEC axial density profile predicted by the GPE with a 10~$\mu$m width Gaussian (to account for finite imaging resolution), and fitting a Gaussian distribution to the resulting profile using nonlinear least-squares.

%----------------------------------------------------------------------------------------------------------------------------
%------------------------------------------------------ Acknowledgements ---------------------------------------------------------
\section*{Acknowledgements}
We thank the Durham soliton theory group for many useful discussions. We acknowledge financial support from the UK Engineering and Physical Sciences Research Council (EPSRC grant EP/F002068/1) and the European Science Foundation within the EUROCORES Programme EuroQUASAR (EPSRC grant EP/G026602/1). TPB was supported by The Marsden Fund of New Zealand (UOO162), and The Royal Society of New Zealand (UOO004).

%%----------------------------------------------------------------------------------------------------------------------------
%%------------------------------------------------------ Author contributions ---------------------------------------------------------
%\section*{Author contributions}
%ALM performed the experiments and data analysis. TPB carried out the numerical simulations. TPW and MMHY assisted with the development of the apparatus. SAG provided theoretical support. SLC conceived and managed the project. ALM, TPB and SLC prepared the manuscript. 
%
%\section*{Competing financial interests}
%The authors declare no competing financial interests.

%----------------------------------------------------------------------------------------------------------------------------
%------------------------------------------------------ Bibliography -------------------------------------------------------

%\bibliography{Bib_Marchant}

\begin{thebibliography}{31}
\expandafter\ifx\csname natexlab\endcsname\relax\def\natexlab#1{#1}\fi
\expandafter\ifx\csname bibnamefont\endcsname\relax
  \def\bibnamefont#1{#1}\fi
\expandafter\ifx\csname bibfnamefont\endcsname\relax
  \def\bibfnamefont#1{#1}\fi
\expandafter\ifx\csname citenamefont\endcsname\relax
  \def\citenamefont#1{#1}\fi
\expandafter\ifx\csname url\endcsname\relax
  \def\url#1{\texttt{#1}}\fi
\expandafter\ifx\csname urlprefix\endcsname\relax\def\urlprefix{URL }\fi
\providecommand{\bibinfo}[2]{#2}
\providecommand{\eprint}[2][]{\url{#2}}

\bibitem[{\citenamefont{Russell}(1844)}]{JSR.solitons}
\bibinfo{author}{\bibfnamefont{J.~S.} \bibnamefont{Russell}}, in
  \emph{\bibinfo{booktitle}{Report of the fourteenth meeting of the British
  association for the advancement of Science}}, edited by
  \bibinfo{editor}{\bibfnamefont{J.}~\bibnamefont{Murray}}
  (\bibinfo{year}{1844}), pp. \bibinfo{pages}{311--90}.

\bibitem[{\citenamefont{Dauxois and Peyrard}(2006)}]{Phys.Solitons}
\bibinfo{author}{\bibfnamefont{T.}~\bibnamefont{Dauxois}} \bibnamefont{and}
  \bibinfo{author}{\bibfnamefont{M.}~\bibnamefont{Peyrard}},
  \emph{\bibinfo{title}{Physics of Solitons}} (\bibinfo{publisher}{Cambridge
  University Press}, \bibinfo{year}{2006}).

\bibitem[{\citenamefont{Cronin et~al.}(2009)\citenamefont{Cronin, Schmiedmayer,
  and Pritchard}}]{RevModPhys.81.1051}
\bibinfo{author}{\bibfnamefont{A.~D.} \bibnamefont{Cronin}},
  \bibinfo{author}{\bibfnamefont{J.}~\bibnamefont{Schmiedmayer}},
  \bibnamefont{and} \bibinfo{author}{\bibfnamefont{D.~E.}
  \bibnamefont{Pritchard}}, \bibinfo{journal}{Rev. Mod. Phys.}
  \textbf{\bibinfo{volume}{81}}, \bibinfo{pages}{1051} (\bibinfo{year}{2009}).

\bibitem[{\citenamefont{Cornish et~al.}(2009)\citenamefont{Cornish, Parker,
  Martin, Judd, Scott, Fromhold, and Adams}}]{CornishPhysicaD}
\bibinfo{author}{\bibfnamefont{S.}~\bibnamefont{Cornish}},
  \bibinfo{author}{\bibfnamefont{N.}~\bibnamefont{Parker}},
  \bibinfo{author}{\bibfnamefont{A.}~\bibnamefont{Martin}},
  \bibinfo{author}{\bibfnamefont{T.}~\bibnamefont{Judd}},
  \bibinfo{author}{\bibfnamefont{R.}~\bibnamefont{Scott}},
  \bibinfo{author}{\bibfnamefont{T.}~\bibnamefont{Fromhold}}, \bibnamefont{and}
  \bibinfo{author}{\bibfnamefont{C.}~\bibnamefont{Adams}},
  \bibinfo{journal}{Physica D: Nonlinear Phenomena}
  \textbf{\bibinfo{volume}{238}}, \bibinfo{pages}{1299 }
  (\bibinfo{year}{2009}), ISSN \bibinfo{issn}{0167-2789}.

\bibitem[{\citenamefont{Weiss and Castin}(2009)}]{PhysRevLett.102.010403}
\bibinfo{author}{\bibfnamefont{C.}~\bibnamefont{Weiss}} \bibnamefont{and}
  \bibinfo{author}{\bibfnamefont{Y.}~\bibnamefont{Castin}},
  \bibinfo{journal}{Phys. Rev. Lett.} \textbf{\bibinfo{volume}{102}},
  \bibinfo{pages}{010403} (\bibinfo{year}{2009}).

\bibitem[{\citenamefont{Streltsov et~al.}(2009)\citenamefont{Streltsov, Alon,
  and Cederbaum}}]{PhysRevA.80.043616}
\bibinfo{author}{\bibfnamefont{A.~I.} \bibnamefont{Streltsov}},
  \bibinfo{author}{\bibfnamefont{O.~E.} \bibnamefont{Alon}}, \bibnamefont{and}
  \bibinfo{author}{\bibfnamefont{L.~S.} \bibnamefont{Cederbaum}},
  \bibinfo{journal}{Phys. Rev. A} \textbf{\bibinfo{volume}{80}},
  \bibinfo{pages}{043616} (\bibinfo{year}{2009}).

\bibitem[{\citenamefont{Pethick and Smith}(2001)}]{PandS}
\bibinfo{author}{\bibfnamefont{C.~J.} \bibnamefont{Pethick}} \bibnamefont{and}
  \bibinfo{author}{\bibfnamefont{H.}~\bibnamefont{Smith}},
  \emph{\bibinfo{title}{Bose-Einstein Condensation in Dilute Gases}}
  (\bibinfo{publisher}{Cambridge University Press}, \bibinfo{year}{2001}).

\bibitem[{\citenamefont{Pitaevskii and Stringari}(2003)}]{PitaevskiiandS}
\bibinfo{author}{\bibfnamefont{L.}~\bibnamefont{Pitaevskii}} \bibnamefont{and}
  \bibinfo{author}{\bibfnamefont{S.}~\bibnamefont{Stringari}},
  \emph{\bibinfo{title}{Bose-Einstein Condensation}}
  (\bibinfo{publisher}{Clarendon Press, Oxford}, \bibinfo{year}{2003}).

\bibitem[{\citenamefont{Billam et~al.}(2012)\citenamefont{Billam, Marchant,
  Cornish, Gardiner, and Parker}}]{Bookchapter}
\bibinfo{author}{\bibfnamefont{T.~P.} \bibnamefont{Billam}},
  \bibinfo{author}{\bibfnamefont{A.~L.} \bibnamefont{Marchant}},
  \bibinfo{author}{\bibfnamefont{S.~L.} \bibnamefont{Cornish}},
  \bibinfo{author}{\bibfnamefont{S.~A.} \bibnamefont{Gardiner}},
  \bibnamefont{and} \bibinfo{author}{\bibfnamefont{N.~G.}
  \bibnamefont{Parker}}, \emph{\bibinfo{title}{Bright solitary matter waves:
  formation, stability and interactions}}
  (\bibinfo{publisher}{arXiv:1209.0560}, \bibinfo{year}{2012}).

\bibitem[{\citenamefont{{Strecker} et~al.}(2002)\citenamefont{{Strecker},
  {Partridge}, {Truscott}, and {Hulet}}}]{2002Natur.417..150S}
\bibinfo{author}{\bibfnamefont{K.~E.} \bibnamefont{{Strecker}}},
  \bibinfo{author}{\bibfnamefont{G.~B.} \bibnamefont{{Partridge}}},
  \bibinfo{author}{\bibfnamefont{A.~G.} \bibnamefont{{Truscott}}},
  \bibnamefont{and} \bibinfo{author}{\bibfnamefont{R.~G.}
  \bibnamefont{{Hulet}}}, \bibinfo{journal}{Nature}
  \textbf{\bibinfo{volume}{417}}, \bibinfo{pages}{150} (\bibinfo{year}{2002}).

\bibitem[{\citenamefont{Khaykovich et~al.}(2002)\citenamefont{Khaykovich,
  Schreck, Ferrari, Bourdel, Cubizolles, Carr, Castin, and
  Salomon}}]{Khaykovich17052002}
\bibinfo{author}{\bibfnamefont{L.}~\bibnamefont{Khaykovich}},
  \bibinfo{author}{\bibfnamefont{F.}~\bibnamefont{Schreck}},
  \bibinfo{author}{\bibfnamefont{G.}~\bibnamefont{Ferrari}},
  \bibinfo{author}{\bibfnamefont{T.}~\bibnamefont{Bourdel}},
  \bibinfo{author}{\bibfnamefont{J.}~\bibnamefont{Cubizolles}},
  \bibinfo{author}{\bibfnamefont{L.~D.} \bibnamefont{Carr}},
  \bibinfo{author}{\bibfnamefont{Y.}~\bibnamefont{Castin}}, \bibnamefont{and}
  \bibinfo{author}{\bibfnamefont{C.}~\bibnamefont{Salomon}},
  \bibinfo{journal}{Science} \textbf{\bibinfo{volume}{296}},
  \bibinfo{pages}{1290} (\bibinfo{year}{2002}).

\bibitem[{\citenamefont{Cornish et~al.}(2006)\citenamefont{Cornish, Thompson,
  and Wieman}}]{PhysRevLett.96.170401}
\bibinfo{author}{\bibfnamefont{S.~L.} \bibnamefont{Cornish}},
  \bibinfo{author}{\bibfnamefont{S.~T.} \bibnamefont{Thompson}},
  \bibnamefont{and} \bibinfo{author}{\bibfnamefont{C.~E.}
  \bibnamefont{Wieman}}, \bibinfo{journal}{Phys. Rev. Lett.}
  \textbf{\bibinfo{volume}{96}}, \bibinfo{pages}{170401}
  (\bibinfo{year}{2006}).

\bibitem[{\citenamefont{Ruprecht et~al.}(1995)\citenamefont{Ruprecht, Holland,
  Burnett, and Edwards}}]{PhysRevA.51.4704}
\bibinfo{author}{\bibfnamefont{P.~A.} \bibnamefont{Ruprecht}},
  \bibinfo{author}{\bibfnamefont{M.~J.} \bibnamefont{Holland}},
  \bibinfo{author}{\bibfnamefont{K.}~\bibnamefont{Burnett}}, \bibnamefont{and}
  \bibinfo{author}{\bibfnamefont{M.}~\bibnamefont{Edwards}},
  \bibinfo{journal}{Phys. Rev. A} \textbf{\bibinfo{volume}{51}},
  \bibinfo{pages}{4704} (\bibinfo{year}{1995}).

\bibitem[{\citenamefont{Dabrowska-W\"uster
  et~al.}(2009)\citenamefont{Dabrowska-W\"uster, W\"uster, and
  Davis}}]{Davis2009}
\bibinfo{author}{\bibfnamefont{B.~J.} \bibnamefont{Dabrowska-W\"uster}},
  \bibinfo{author}{\bibfnamefont{S.}~\bibnamefont{W\"uster}}, \bibnamefont{and}
  \bibinfo{author}{\bibfnamefont{M.~J.} \bibnamefont{Davis}},
  \bibinfo{journal}{New Journal of Physics} \textbf{\bibinfo{volume}{11}},
  \bibinfo{pages}{053017} (\bibinfo{year}{2009}).

\bibitem[{\citenamefont{Streltsov et~al.}(2008)\citenamefont{Streltsov, Alon,
  and Cederbaum}}]{PhysRevLett.100.130401}
\bibinfo{author}{\bibfnamefont{A.~I.} \bibnamefont{Streltsov}},
  \bibinfo{author}{\bibfnamefont{O.~E.} \bibnamefont{Alon}}, \bibnamefont{and}
  \bibinfo{author}{\bibfnamefont{L.~S.} \bibnamefont{Cederbaum}},
  \bibinfo{journal}{Phys. Rev. Lett.} \textbf{\bibinfo{volume}{100}},
  \bibinfo{pages}{130401} (\bibinfo{year}{2008}).

\bibitem[{\citenamefont{Helm et~al.}(2012)\citenamefont{Helm, Billam, and
  Gardiner}}]{PhysRevA.85.053621}
\bibinfo{author}{\bibfnamefont{J.~L.} \bibnamefont{Helm}},
  \bibinfo{author}{\bibfnamefont{T.~P.} \bibnamefont{Billam}},
  \bibnamefont{and} \bibinfo{author}{\bibfnamefont{S.~A.}
  \bibnamefont{Gardiner}}, \bibinfo{journal}{Phys. Rev. A}
  \textbf{\bibinfo{volume}{85}}, \bibinfo{pages}{053621}
  (\bibinfo{year}{2012}).

\bibitem[{\citenamefont{Billam et~al.}(2011)\citenamefont{Billam, Cornish, and
  Gardiner}}]{PhysRevA.83.041602}
\bibinfo{author}{\bibfnamefont{T.~P.} \bibnamefont{Billam}},
  \bibinfo{author}{\bibfnamefont{S.~L.} \bibnamefont{Cornish}},
  \bibnamefont{and} \bibinfo{author}{\bibfnamefont{S.~A.}
  \bibnamefont{Gardiner}}, \bibinfo{journal}{Phys. Rev. A}
  \textbf{\bibinfo{volume}{83}}, \bibinfo{pages}{041602}
  (\bibinfo{year}{2011}).

\bibitem[{\citenamefont{Martin and Ruostekoski}(2012)}]{1367-2630-14-4-043040}
\bibinfo{author}{\bibfnamefont{A.~D.} \bibnamefont{Martin}} \bibnamefont{and}
  \bibinfo{author}{\bibfnamefont{J.}~\bibnamefont{Ruostekoski}},
  \bibinfo{journal}{New Journal of Physics} \textbf{\bibinfo{volume}{14}},
  \bibinfo{pages}{043040} (\bibinfo{year}{2012}).

\bibitem[{\citenamefont{Ernst and Brand}(2010)}]{PhysRevA.81.033614}
\bibinfo{author}{\bibfnamefont{T.}~\bibnamefont{Ernst}} \bibnamefont{and}
  \bibinfo{author}{\bibfnamefont{J.}~\bibnamefont{Brand}},
  \bibinfo{journal}{Phys. Rev. A} \textbf{\bibinfo{volume}{81}},
  \bibinfo{pages}{033614} (\bibinfo{year}{2010}).

\bibitem[{\citenamefont{Cuevas et~al.}(2013)\citenamefont{Cuevas, Kevrekidis,
  Malomed, Dyke, and Hulet}}]{Hulet_theory}
\bibinfo{author}{\bibfnamefont{J.}~\bibnamefont{Cuevas}},
  \bibinfo{author}{\bibfnamefont{P.~G.} \bibnamefont{Kevrekidis}},
  \bibinfo{author}{\bibfnamefont{B.~A.} \bibnamefont{Malomed}},
  \bibinfo{author}{\bibfnamefont{P.}~\bibnamefont{Dyke}}, \bibnamefont{and}
  \bibinfo{author}{\bibfnamefont{R.~G.} \bibnamefont{Hulet}},
  \bibinfo{journal}{arXiv:1301.3959}  (\bibinfo{year}{2013}).

\bibitem[{\citenamefont{Dyke et~al.}(2011)\citenamefont{Dyke, Sidong, Pollack,
  Dries, and Hulet}}]{Huletexpt}
\bibinfo{author}{\bibfnamefont{P.}~\bibnamefont{Dyke}},
  \bibinfo{author}{\bibfnamefont{L.}~\bibnamefont{Sidong}},
  \bibinfo{author}{\bibfnamefont{S.}~\bibnamefont{Pollack}},
  \bibinfo{author}{\bibfnamefont{D.}~\bibnamefont{Dries}}, \bibnamefont{and}
  \bibinfo{author}{\bibfnamefont{R.}~\bibnamefont{Hulet}}, in
  \emph{\bibinfo{booktitle}{42nd Annual Meeting of the APS Division of Atomic,
  Molecular and Optical Physics}} (\bibinfo{year}{2011}), vol.
  \bibinfo{volume}{56, No. 5}.

\bibitem[{\citenamefont{Gertjerenken et~al.}(2012)\citenamefont{Gertjerenken,
  Billam, Khaykovich, and Weiss}}]{PhysRevA.86.033608}
\bibinfo{author}{\bibfnamefont{B.}~\bibnamefont{Gertjerenken}},
  \bibinfo{author}{\bibfnamefont{T.~P.} \bibnamefont{Billam}},
  \bibinfo{author}{\bibfnamefont{L.}~\bibnamefont{Khaykovich}},
  \bibnamefont{and} \bibinfo{author}{\bibfnamefont{C.}~\bibnamefont{Weiss}},
  \bibinfo{journal}{Phys. Rev. A} \textbf{\bibinfo{volume}{86}},
  \bibinfo{pages}{033608} (\bibinfo{year}{2012}).

\bibitem[{\citenamefont{Lieb and Liniger}(1963)}]{PhysRev.130.1605}
\bibinfo{author}{\bibfnamefont{E.~H.} \bibnamefont{Lieb}} \bibnamefont{and}
  \bibinfo{author}{\bibfnamefont{W.}~\bibnamefont{Liniger}},
  \bibinfo{journal}{Phys. Rev.} \textbf{\bibinfo{volume}{130}},
  \bibinfo{pages}{1605} (\bibinfo{year}{1963}).

\bibitem[{\citenamefont{Al~Khawaja et~al.}(2002)\citenamefont{Al~Khawaja,
  Stoof, Hulet, Strecker, and Partridge}}]{PhysRevLett.89.200404}
\bibinfo{author}{\bibfnamefont{U.}~\bibnamefont{Al~Khawaja}},
  \bibinfo{author}{\bibfnamefont{H.~T.~C.} \bibnamefont{Stoof}},
  \bibinfo{author}{\bibfnamefont{R.~G.} \bibnamefont{Hulet}},
  \bibinfo{author}{\bibfnamefont{K.~E.} \bibnamefont{Strecker}},
  \bibnamefont{and} \bibinfo{author}{\bibfnamefont{G.~B.}
  \bibnamefont{Partridge}}, \bibinfo{journal}{Phys. Rev. Lett.}
  \textbf{\bibinfo{volume}{89}}, \bibinfo{pages}{200404}
  (\bibinfo{year}{2002}).

\bibitem[{\citenamefont{Parker et~al.}(2009)\citenamefont{Parker, Martin,
  Adams, and Cornish}}]{Parker20091456}
\bibinfo{author}{\bibfnamefont{N.}~\bibnamefont{Parker}},
  \bibinfo{author}{\bibfnamefont{A.}~\bibnamefont{Martin}},
  \bibinfo{author}{\bibfnamefont{C.}~\bibnamefont{Adams}}, \bibnamefont{and}
  \bibinfo{author}{\bibfnamefont{S.}~\bibnamefont{Cornish}},
  \bibinfo{journal}{Physica D: Nonlinear Phenomena}
  \textbf{\bibinfo{volume}{238}}, \bibinfo{pages}{1456 }
  (\bibinfo{year}{2009}), ISSN \bibinfo{issn}{0167-2789}.

\bibitem[{\citenamefont{Parker et~al.}(2008)\citenamefont{Parker, Martin,
  Cornish, and Adams}}]{Parker.JPhysB}
\bibinfo{author}{\bibfnamefont{N.~G.} \bibnamefont{Parker}},
  \bibinfo{author}{\bibfnamefont{A.~M.} \bibnamefont{Martin}},
  \bibinfo{author}{\bibfnamefont{S.~L.} \bibnamefont{Cornish}},
  \bibnamefont{and} \bibinfo{author}{\bibfnamefont{C.~S.} \bibnamefont{Adams}},
  \bibinfo{journal}{Journal of Physics B: Atomic, Molecular and Optical
  Physics} \textbf{\bibinfo{volume}{41}}, \bibinfo{pages}{045303}
  (\bibinfo{year}{2008}).

\bibitem[{\citenamefont{Carr and Brand}(2004)}]{PhysRevLett.92.040401}
\bibinfo{author}{\bibfnamefont{L.~D.} \bibnamefont{Carr}} \bibnamefont{and}
  \bibinfo{author}{\bibfnamefont{J.}~\bibnamefont{Brand}},
  \bibinfo{journal}{Phys. Rev. Lett.} \textbf{\bibinfo{volume}{92}},
  \bibinfo{pages}{040401} (\bibinfo{year}{2004}).

\bibitem[{\citenamefont{Dimopoulos and Geraci}(2003)}]{PhysRevD.68.124021}
\bibinfo{author}{\bibfnamefont{S.}~\bibnamefont{Dimopoulos}} \bibnamefont{and}
  \bibinfo{author}{\bibfnamefont{A.~A.} \bibnamefont{Geraci}},
  \bibinfo{journal}{Phys. Rev. D} \textbf{\bibinfo{volume}{68}},
  \bibinfo{pages}{124021} (\bibinfo{year}{2003}).

\bibitem[{\citenamefont{Marchant et~al.}(2012)\citenamefont{Marchant, H\"andel,
  Hopkins, Wiles, and Cornish}}]{PhysRevA.85.053647}
\bibinfo{author}{\bibfnamefont{A.~L.} \bibnamefont{Marchant}},
  \bibinfo{author}{\bibfnamefont{S.}~\bibnamefont{H\"andel}},
  \bibinfo{author}{\bibfnamefont{S.~A.} \bibnamefont{Hopkins}},
  \bibinfo{author}{\bibfnamefont{T.~P.} \bibnamefont{Wiles}}, \bibnamefont{and}
  \bibinfo{author}{\bibfnamefont{S.~L.} \bibnamefont{Cornish}},
  \bibinfo{journal}{Phys. Rev. A} \textbf{\bibinfo{volume}{85}},
  \bibinfo{pages}{053647} (\bibinfo{year}{2012}).

\bibitem[{\citenamefont{Lin et~al.}(2009)\citenamefont{Lin, Perry, Compton,
  Spielman, and Porto}}]{PhysRevA.79.063631}
\bibinfo{author}{\bibfnamefont{Y.-J.} \bibnamefont{Lin}},
  \bibinfo{author}{\bibfnamefont{A.~R.} \bibnamefont{Perry}},
  \bibinfo{author}{\bibfnamefont{R.~L.} \bibnamefont{Compton}},
  \bibinfo{author}{\bibfnamefont{I.~B.} \bibnamefont{Spielman}},
  \bibnamefont{and} \bibinfo{author}{\bibfnamefont{J.~V.} \bibnamefont{Porto}},
  \bibinfo{journal}{Phys. Rev. A} \textbf{\bibinfo{volume}{79}},
  \bibinfo{pages}{063631} (\bibinfo{year}{2009}).

\bibitem[{\citenamefont{Marchant et~al.}(2011)\citenamefont{Marchant, H\"andel,
  Wiles, Hopkins, and Cornish}}]{1367-2630-13-12-125003}
\bibinfo{author}{\bibfnamefont{A.~L.} \bibnamefont{Marchant}},
  \bibinfo{author}{\bibfnamefont{S.}~\bibnamefont{H\"andel}},
  \bibinfo{author}{\bibfnamefont{T.~P.} \bibnamefont{Wiles}},
  \bibinfo{author}{\bibfnamefont{S.~A.} \bibnamefont{Hopkins}},
  \bibnamefont{and} \bibinfo{author}{\bibfnamefont{S.~L.}
  \bibnamefont{Cornish}}, \bibinfo{journal}{New Journal of Physics}
  \textbf{\bibinfo{volume}{13}}, \bibinfo{pages}{125003}
  (\bibinfo{year}{2011}).

\end{thebibliography}

\end{document}